\documentclass[twocolumn]{aastex631}

\usepackage{graphicx} 
\usepackage{apjfonts} 
\usepackage{hyperref}
\usepackage{subfigure}
\usepackage{xcolor}

\shorttitle{The solar internetwork. III. Unipolar versus bipolar flux appearance} 

\shortauthors{Go\v{s}i\'{c} et al.}

\begin{document}

\title{The solar internetwork. III. Unipolar versus bipolar flux appearance}

\correspondingauthor{M.~Go\v{s}i\'{c}}
\email{gosic@baeri.org}

\author[0000-0002-5879-4371]{M.~Go\v{s}i\'{c}}
\affil{Instituto de Astrof\'{\i}sica de Andaluc\'{\i}a (IAA-CSIC), Apdo.\ 3004, 18080 Granada, Spain} 
\affil{Lockheed Martin Solar and Astrophysics Laboratory, Palo Alto, CA 94304, USA}
\affil{Bay Area Environmental Research Institute, Moffett Field, CA 94035, USA}

\author[0000-0001-8669-8857]{L.~R.~Bellot Rubio}
\affil{Instituto de Astrof\'{\i}sica de Andaluc\'{\i}a (IAA-CSIC), Apdo.\ 3004, 18080 Granada, Spain} 

\author[0000-0003- 2110-9753]{M.~C.~M.~Cheung}
\affil{Lockheed Martin Solar and Astrophysics Laboratory, Palo Alto, CA 94304, USA}

\author[0000-0001-8829-1938]{D.~Orozco Su\'arez}
\affil{Instituto de Astrof\'{\i}sica de Andaluc\'{\i}a (IAA-CSIC), Apdo.\ 3004, 18080 Granada, Spain} 

\author[0000-0002-5054-8782]{Y.~Katsukawa}
\affil{National Astronomical Observatory of Japan, 2-21-1 Osawa, Mitaka, Tokyo 181-8588, Japan}

\author[0000-0002-3387-026X]{J.~C.~del Toro Iniesta}
\affil{Instituto de Astrof\'{\i}sica de Andaluc\'{\i}a (IAA-CSIC), Apdo.\ 3004, 18080 Granada, Spain} 

\begin{abstract}

Small-scale internetwork (IN) magnetic fields are considered to be the
main building blocks of the quiet Sun magnetism. For this reason, it
is crucial to understand how they appear on the solar surface. Here,
we employ a high-resolution, high-sensitivity, long-duration
Hinode/NFI magnetogram sequence to analyze the appearance modes and
spatio-temporal evolution of individual IN magnetic elements inside a
supergranular cell at the disk center. From identification of flux
patches and magnetofrictional simulations, we show that there are two
distinct populations of IN flux concentrations: unipolar and bipolar
features. Bipolar features tend to be bigger and stronger than
unipolar features. They also live longer and carry more flux per
feature. Both types of flux concentrations appear uniformly over the
solar surface. However, we argue that bipolar features truly represent
the emergence of new flux on the solar surface, while unipolar
features seem to be formed by coalescence of background flux. Magnetic
bipoles appear at a faster rate than unipolar features (68 as opposed
to 55~Mx~cm$^{-2}$~day$^{-1}$), and provide about $70$\% of the total
instantaneous IN flux detected in the interior of the supergranule.

\end{abstract}


\section{Introduction}

The Sun was perceived for a long time to be magnetic only in the
regions occupied by sunspots, the so-called active regions. Everywhere
else, the solar surface seemed to be devoid of magnetic activity, and
was consequently denoted as the quiet Sun (QS). Such an impression
remained intact until the 1970s, when the network was discovered
\citep{Sheeley1969} and weak small-scale magnetic features were seen
in the interior of supergranular cells \citep{LivingstonHarvey1971,
LivingstonHarvey, Smithson1975}. Since then, magnetic fields have been
observed everywhere on the quiet solar surface. At the boundaries of
supergranular cells, strong kG fields form the photospheric network
(NE), while weak and highly transient internetwork (IN) fields
pervade the interior of supergranules.

IN fields are believed to be an essential contributor to the magnetic
flux and energy budget of the solar photosphere
\citep[e.g.,][]{TrujilloBueno2004}. Indeed, measurements from
different instruments have revealed that up to 50\% of the total 
QS flux is in the form of small, weak IN flux patches
\citep{Wangetal1995, Meunieretal1998, Lites2002, Zhouetal2013,
Gosic2014}. IN magnetic elements bring flux to the solar surface at a
rate of $120$~Mx~cm$^{-2}$~day$^{-1}$ \citep{Gosic2016}, much faster
than active regions \cite[0.1~Mx~cm$^{-2}$~day$^{-1}$;][]{SchrijverHarvey}.  
Observations at 0\farcs1 resolution suggest even larger rates of
$1100$~Mx~cm$^{-2}$~day$^{-1}$ \citep{Smithaetal2017}. Such a
tremendous inflow of IN flux is capable of maintaining the
photospheric NE \citep{Gosic2014, Giannattasioetal2020}. These
findings establish IN flux features as one of the main contributors to
the entire QS magnetic flux, which has an important consequence---to
understand the magnetism of our star we need to understand how they are
formed and how they evolve.

IN magnetic features are observed to appear on the solar surface as
small bipolar flux concentrations called magnetic loops
\citep{MartinezGonzalezetal2007, Centenoetal2007,
MartinezGonzalezBellotRubio2009, Gomory2010, Guglielminoetal2012,
Palaciosetal2012}. They first show linear polarization at photospheric
levels, revealing the horizontal fields of the loop tops. This
is followed by positive and negative circular polarization patches
flanking the linear polarization patch, which correspond to more
vertical fields (the loop footpoints). As the loop rises to higher
layers, the linear signals disappear and the vertical fields remain
visible, drifting away from each other. \cite{Fischeretal2020} have
described similar bipolar features emerging in granular lanes, perhaps
as a consequence of shallow recirculation of magnetic flux by granular
vortex flows.

Magnetic features inside IN regions are also observed to appear as
isolated unipolar flux patches within intergranular lanes
\citep[e.g.,][]{Martin1988, Lamb2008, Lamb2010} or above granules
\citep{Orozco2008}. No opposite polarity patches can be detected nearby, 
although this does not necessarily mean they are not present on the
surface.

About $8\%$ of the IN flux concentrations observed 
in the photosphere may be the result of preexisting magnetic fields being
dragged down from the canopy that overlies the internetwork.  Patches
created through this mechanism could appear in any of the two forms,
as unipolar or bipolar features, and would be embedded in downflows
\citep{Danilovicetal2010b, Pietarilaetal2011}.

Using observations from the Michelson Doppler Imager on the Solar and
Heliospheric Observatory \citep{Scherrer}, \cite{Lamb2008} estimated
that $94$\% of the features containing new flux are unipolar at a
resolution of $1\farcs2$. According to \cite{Lamb2008}, mergings and
fragmentations of flux features merely rearrange the flux, i.e., they
do not bring new flux to the solar surface. Taking this into account,
the results published by \cite{Anusha2017} in their Table 2 for the
10:1 area-ratio criterion suggest that 8728 out of 9093 birth events
bringing new flux to the surface, i.e., 96\% of the features, are
unipolar in magnetograms obtained with the Imaging Magnetograph
eXperiment \citep[IMaX;][]{MartinezPilletetal2011} on board the
SUNRISE observatory \citep{Solankietal2010, Bartholetal2011}. Also,
from the results of \cite{Smithaetal2017} it follows that 91\% of the
newly appeared flux is due to unipolar structures, under the
assumption that merging and fragmentation processes are as frequent
for unipolar features as for bipolar features (H. N. Smitha 2021,
private communication). Having such a large fraction of the total IN
flux in unipolar form is not compatible with Maxwell equations, as the
divergence of the magnetic field must be zero. To solve this problem
it is necessary to understand where the missing opposite polarity flux
is located. Well established theoretical models produce bipolar
features, rather than unipolar features. In those models IN fields are
generated by small-scale surface dynamo action (\citealt{Cattaneo1999,
CattaneoHughes2001, VoglerSchussler2007, Danilovicetal2010,
Rempel2014}), flux recycling from decayed active regions
\citep[e.g.,][]{Ploneretal2001, DeWijnetal2005}, flux emergence
from subphotospheric layers \citep{DeWijnetal2009} similar to
ephemeral regions \citep{HarveyMartin1973, Harveyetal1975,
Hagenaar2001}, and shallow recirculation in granular convection 
 \citep{Rempel2018, Fischeretal2020}.

The fraction of flux that appears on the solar surface in bipolar form
is difficult to quantify because the poles of magnetic loops must be
detected first. Unfortunately, from individual longitudinal
magnetograms one can never be sure if two or more opposite-polarity
patches that lie relatively close to each other are part of the same
loop or not. Thus, to identify small-scale bipolar features in IN
regions, it is necessary to study the temporal evolution of the flux
patches and the magnetic connectivity between them. This is feasible
only by using space-borne observations that allow us to examine the
continuous evolution of IN magnetic features on temporal scales from
minutes to hours at the highest resolution and sensitivity possible.

The work presented here is an attempt to distinguish unipolar and
bipolar features in IN regions. We study how and where IN features
appear, how they evolve with time, and to what extent they contribute
to the total IN flux budget and flux appearance rate. For the first
time, we are able to examine the properties of unipolar and bipolar
features separately. To study the connectivity between
magnetic patches we use the magnetofrictional method \citep{Yang,
Craig}. We prefer this method over potential field extrapolations
because it accounts for the history of the features and QS magnetic
fields may have a significant non-potential component
\citep{WoodardChae1999, Zhaoetal2009}.

The paper is organized as follows. Section~\ref{sect2} describes the
observations. Section~\ref{sect3} explains the methods we use to track
individual IN flux patches and to identify bipolar features among
them. In Section~\ref{sect4} we present the properties of unipolar 
and bipolar IN features and study their spatial distribution, their
contribution to the total instantaneous IN flux, and their flux
appearance rate. Finally, Section~\ref{sect5} summarizes our findings
and conclusions.

\begin{figure*}[!t]
	\begin{center}
		\resizebox{1\hsize}{!}{\includegraphics[bb=20 0 907 283]{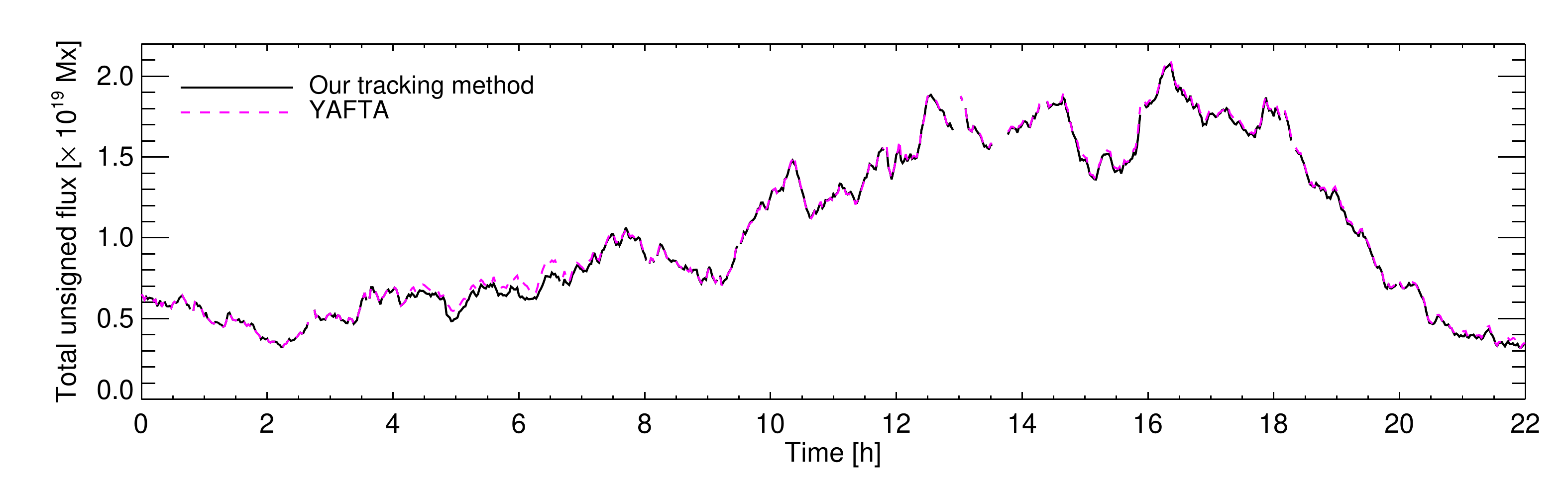}} 
		\caption{Temporal evolution of the total unsigned flux detected in the selected circular region. The solid black line represents the total IN flux obtained using our tracking method, while the dashed violet line gives the total flux obtained with YAFTA. $t$=0 corresponds to 08:31 UT November 2, 2010.}
		\label{fig1}
	\end{center}
\end{figure*}

\section{Observations and data processing}
\label{sect2}

The data set used in this work was acquired on 2010 November 2--3 with the
Narrowband Filter Imager \citep[NFI;][]{Tsuneta} aboard the Hinode
satellite \citep{2007SoPh..243....3K}. The measurements belong to
the Hinode Operation Plan 151 and are described in detail by
\citet{Gosic2014}. Here we briefly summarize the main
characteristics of this observational sequence. The NFI was operated
in shutterless mode to achieve the highest possible sensitivity. We
took Stokes \textit{I} and \textit{V} filtergrams in the \ion{Na}{1}
589.6~nm line at two wavelength positions ($\pm 16$~pm from the line
center) that sample the mid-upper photosphere. They were used to
calculate magnetograms and Dopplergrams. The effective exposure time
of the magnetograms was 6.4~s. This resulted in a noise level of
6~Mx~cm$^{-2}$, which was further reduced to 4~Mx~cm$^{-2}$ applying a
$3\times3$ Gaussian-type spatial kernel. The five-minute photospheric
oscillations were removed from the magnetograms and Dopplergrams using
a subsonic filter \citep{1989ApJ...336..475T, 1992AA...256..652S}.

Our observations show the temporal evolution of a supergranular cell
located at the center of the solar disk from its early formation phase
until maturity. This cell is highly unipolar, with negative patches
dominating over positive patches. Indeed, about 85\% of its total
unsigned network flux of $\sim3.5\times10^{20}$~Mx is negative. The
entire supergranule (effective radius $\sim$13~Mm) is visible within
the field of view (FOV) of $80\arcsec \times 74\arcsec$ during
38~hours of observations. However, since the measurements have two
short gaps due to telemetry problems, the longest sequence without
interruptions lasts $\sim24$ hours, of which we used 22 hours to study
the complete history of IN features (from appearance to
disappearance). This data set is ideal for investigating the
spatio-temporal evolution of IN magnetic features due to its high
cadence of 90~s and spatial resolution of about $0\farcs32$ (pixel
size of $0\farcs16$).

\section{Methods}
\label{sect3}

To study how IN magnetic features appear on the solar surface, we have
to identify each individual flux patch as soon as it becomes visible
in the magnetograms and determine whether it is unipolar or
bipolar. Since IN features can interact with other features, we also
have to track their temporal evolution and detect all merging and
fragmentation processes they undergo during their lifetimes. Correct
identification of merging and fragmentation events is crucial for a
reliable calculation of the flux content of individual
features. Below, we describe each of these steps in detail.

\subsection{Detection and identification of magnetic features}

We have automatically tracked all magnetic features visible in the
magnetogram sequence within a circle of $9.3$~Mm radius, whose center
always coincides with the center of the examined supergranular
cell. The circular area follows the temporal evolution of the
supergranule. For the purpose of detection of magnetic features, we
set a flux density threshold of $3\sigma$, i.e., 12~Mx~cm$^{-2}$. As
additional requirements, we use a minimum size of 4 pixels and a
minimum lifetime of two frames (1.5 minutes). We consider a
feature to live from the frame in which it becomes visible for the
first time (through in-situ appearance or fragmentation) until the
moment it disappears (through fading, cancellation or merging with a
stronger flux patch).

The identification of features is done automatically using the
clumping method \citep{Parnell2009} or the downhill method
\citep{WelschLongcope}, with manual verification and correction as
needed. The clumping method groups into one patch all contiguous
pixels above the threshold that have the same sign. This is the
default for identification. We use the downhill method only when
magnetic features start to merge so that the interacting features can
be identified for as long as possible. Each newly detected magnetic
feature receives a unique label and thus can be accessed at any time.

To make sure that our method is reliable, we compared the temporal
evolution of the detected unsigned flux with the results of a YAFTA
run \citep[]{WelschLongcope}. This is shown in Figure~\ref{fig1}. As
can be seen, the total instantaneous fluxes in the selected circular
region match almost perfectly. Small differences are the consequence
of the different identification methods used by our tracking code
(clumping and downhill) and YAFTA (only clumping). Also, YAFTA may
capture more network pixels in the first several hours of the tracking
when the selected circular region is closer to the edges of the
supergranular cell. All this may explain the slightly higher fluxes
measured by YAFTA in some frames.

\subsection{Identification of magnetic loops and clusters}
\label{sect33}

Magnetic flux appears on the solar surface as unipolar or bipolar
features. The latter include magnetic loops and clusters. This means
there are three distinct groups of flux patches inside supergranules:

\begin{enumerate}	
\item Unipolar patches appearing as isolated flux concentrations;
\item Loop footpoints observed as positive and negative circular polarization
patches moving away from each other;
\item Flux clusters, i.e., structures made up of several bipolar patches 
that emerge within a short time interval in a relatively small area.
\end{enumerate}

Detecting magnetic loops without the help of linear polarization
measurements is not trivial because the loop tops cannot be
observed. One has to rely on circular polarization measurements to
locate the loop footpoints. However, the information they
provide is often insufficient to decide if two or more
opposite-polarity patches close in space and time are truly the
footpoints of a loop (hence bipolar) or unrelated patches (hence
unipolar). To minimize this problem we also use the intensity maps and
Dopplergrams derived from the NFI observations. They tell us whether
the patches appear above granules, at their edges, or in intergranular
lanes. In addition, we examine the magnetic connectivity between flux
patches using a magnetofrictional simulation of the data. 

Thus, the following criteria are considered to identify magnetic loops inside the supergranular cell:

\begin{enumerate}
	\item Type of features. We search for loop footpoints among all the
	flux patches detected to appear in situ in the selected region of the
	supergranular cell.

	\item Timing of footpoint appearance. To consider a flux patch
	as a possible loop footpoint, it has to appear in the vicinity
	of an opposite-polarity flux concentration emerging at
	most 6 minutes (5 frames) earlier or
	later\footnote{For $60$\% of the detected magnetic
	loops, the two footpoints appear in the same or the next
	frame. The percentage increases to over $80$\% and $90$\% when
	$3$ and $4$ frames are considered, respectively. This suggests
	that $5$ frames is a reasonable value for the maximum time
	lag. Increasing it further would provide negligible benefit
	in the detection of bipoles.}.

	\item Flux content of footpoints. The total unsigned fluxes of the
	possible footpoints of a magnetic loop cannot differ by more than a
	factor of 3. This parameter is based on our previous knowledge 
	of the properties of IN magnetic features \citep{Gosic2014, Gosic2016}.

	\item Separation of footpoints. The patches must move in opposite
	directions, following a more or less straight trajectory.

	\item Site of emergence. The two footpoints of a loop must emerge in
	or at the edges of the same granule. We use the intensity
	filtergrams and Dopplergrams to find the position of the patches.

	\item Footpoint connectivity. The two opposite-polarity
	footpoints should be magnetically connected, according to the
	magnetofrictional simulation.
\end{enumerate}

The last criterion makes it possible to identify magnetic loops with
confidence, but it is sometimes waived. The reason is that about 25\%
of the footpoints detected in the cell are too weak to be properly
modeled by the magnetofrictional method. Magnetogram signals close to
the noise level introduce large uncertainties in the calculation of
the magnetic flux, the local flux balance, and the velocity and
electric field, which may affect the extrapolations, and therefore
the connectivity between magnetic features. In addition, when very
weak footpoints are located close to strong flux features, they may
incorrectly be connected to those features rather than to each
other. This situation is more likely to happen when newly emerging
footpoints appear and immediately start to interact with strong
preexisting patches in their vicinity.

The identification of flux patches  belonging to clusters of
mixed-polarity features is carried out in a somewhat different
manner. The reason is that those patches appear in more populated regions
where the frequency of surface processes is increased. Therefore, we
cannot search for the same number of positive and negative patches or
take into account their flux content. To qualify as cluster members,
magnetic patches must (a) appear in situ within a group of
mixed-polarity features; (b) emerge more or less at the same time
in a relatively small region; and (c) move outward in opposite
directions. The majority of flux patches appearing within clusters
are magnetically connected based on the magnetofrictional simulation.

\subsection{Magnetofrictional simulation}

The magnetic connectivity between patches (or lack thereof) is
determined by tracing magnetic field lines obtained from a data-driven
magnetofrictional simulation of the Hinode/NFI circular polarization
measurements. The magnetofrictional method makes it possible to
construct magnetic field models evolving with time. The method is
based on the assumption that the plasma velocity $\mathbf{v}$ is
proportional to the Lorentz force, i.e.,

\begin{equation}
\mathbf{v}=\frac{1}{\nu} \, \mathbf{j}\times\mathbf{B}.
\end{equation}
Here, $\nu$ represents the frictional coefficient and $\mathbf{j}=\nabla\times\mathbf{B}$ the current density. The evolution of the magnetic field is obtained according to the induction equation
\begin{equation}
\frac{\partial\mathbf{B}}{\partial 
	t}=\nabla\times(\mathbf{v}\times\mathbf{B}).
\end{equation}

In the magnetofrictional code used here, developed by \citeauthor{Mark2012} (\citeyear{Mark2012}; see also \citeauthor{Mark2015} \citeyear{Mark2015}), the induction equation is solved for the vector potential $\mathbf{A}$, i.e.,
\begin{equation}
\frac{\partial\mathbf{A}}{\partial 
	t}=\mathbf{v}\times\mathbf{B}.
\end{equation}
The vector potential is defined by the condition
$\mathbf{B}=\nabla\times\mathbf{A}$.

Assuming that the magnetic field is potential and periodic in the $x$ and $y$ directions, the vector potential can be derived from the longitudinal component of the magnetic field, $B_{z}$. \cite{Chaeetal2001f} showed that the Fourier solution for the vector potential in Cartesian coordinates is given by
\begin{eqnarray}
A_{x} &=& FT^{-1}\left[\frac{ik_{y}}{k_{x}^{2}+k_{y}^{2}}FT(B_{z})\right], \\
A_{y} &=& FT^{-1}\left[-\frac{ik_{x}}{k_{x}^{2}+k_{y}^{2}}FT(B_{z})\right].
\end{eqnarray}
Here, $FT$ represents the Fourier transform and $FT^{-1}$ its inverse. The vector potential of the first magnetogram in the sequence is used as the initial condition of the simulation.

\begin{figure}[t]
	\begin{center}
		\resizebox{1\hsize}{!}{\includegraphics[width=0.99\textwidth, bb=0 0 485 443]{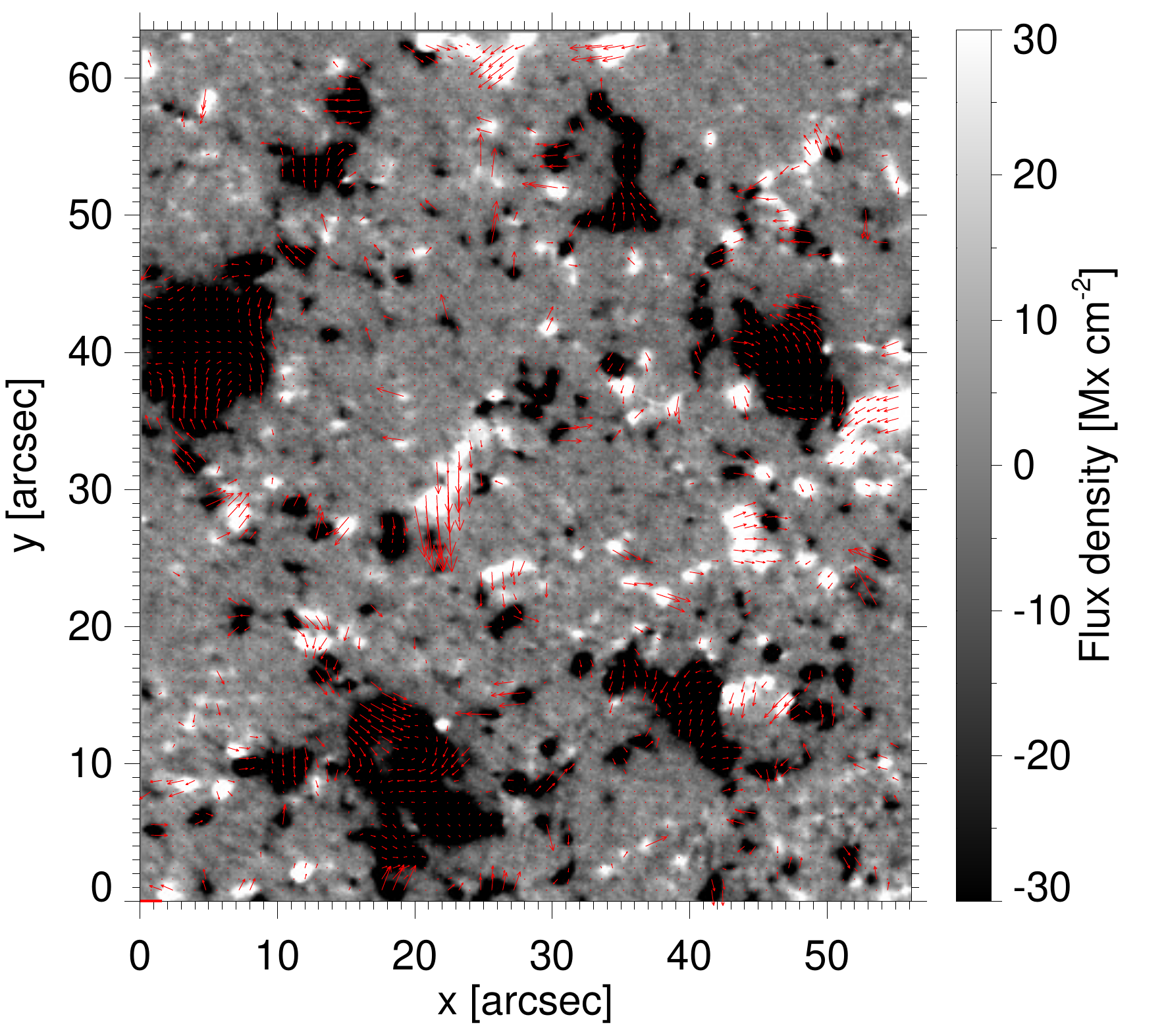}} 
		\caption{Example of a horizontal velocity map. The background image is the
			average of the magnetograms used to calculate the horizontal velocities.}
        \vspace*{-1em}
		\label{fig2}
	\end{center}
\end{figure}

\begin{figure}[t]
	\centering
	\resizebox{1\hsize}{!}{\includegraphics[width=0.99\textwidth]{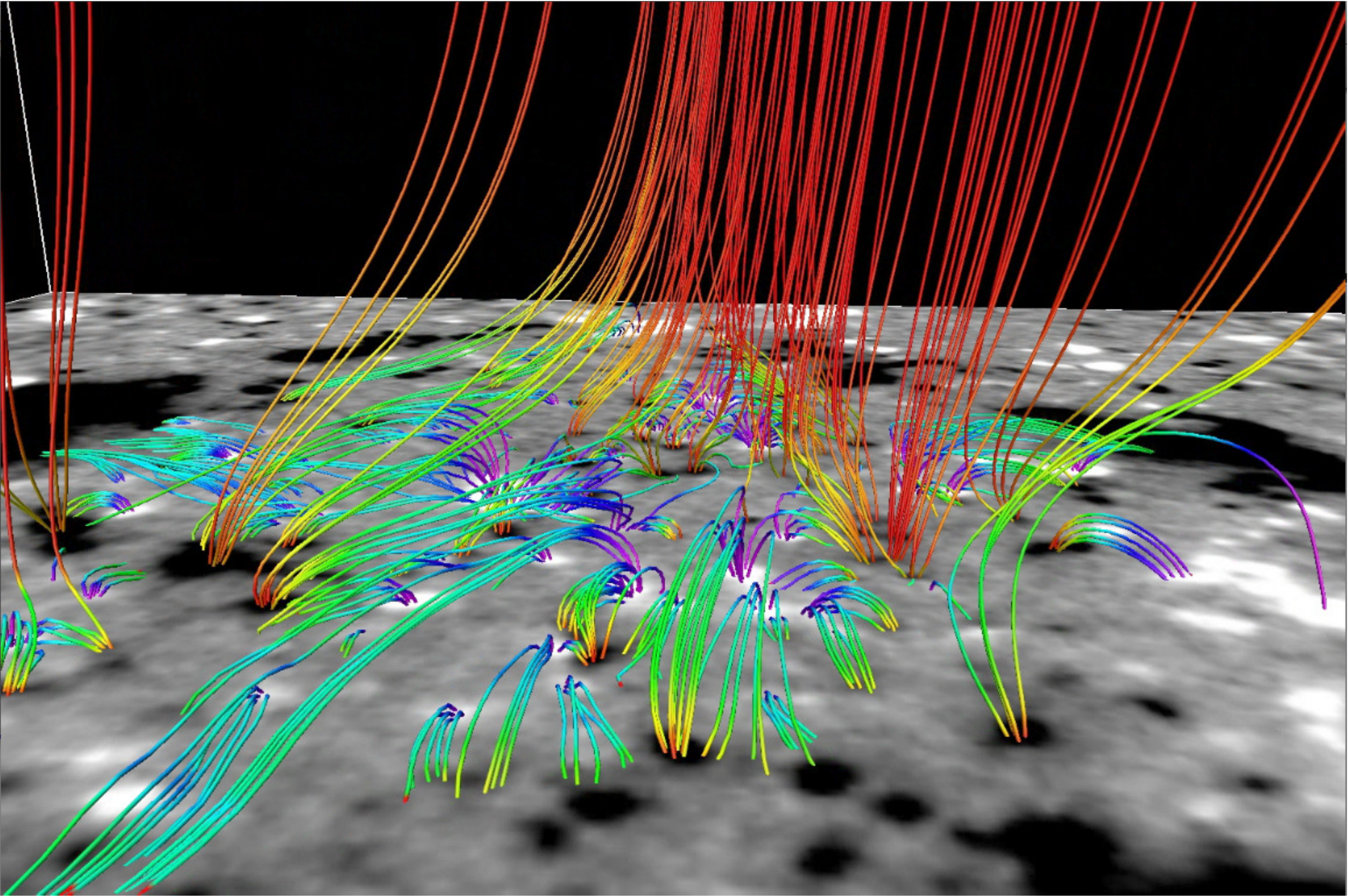}}
	\caption{Snapshot from a magnetofrictional simulation of the
	NFI data set. Field lines are plotted over the corresponding
	magnetogram saturated at $\pm30$~Mx~cm$^{-2}$. The different
	colors indicate the cosine of the magnetic field inclination,
	from $+1$ (violet, positive polarity) to $-1$ (red, negative
	polarity). An animation of this figure showing the
	spatio-temporal evolution of the detected flux features is
	available in the online journal. The animation covers $\sim3$
	hours of solar time from November 2, 2010 at 20:01 UT (12
	seconds real time). The 3D rendering was created using VAPOR
	\citep{Lietal2019}.}  
	\label{fig3}
\end{figure}

\begin{figure*}[!t]
	\centering
	\resizebox{1\hsize}{!}{\includegraphics[width=0.99\textwidth]{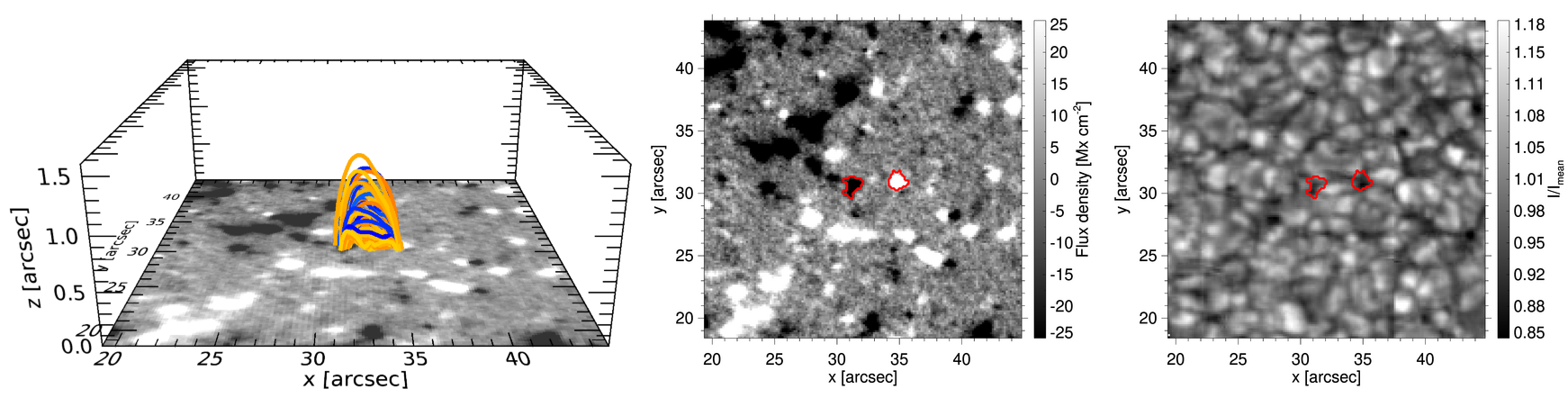}}
	\caption{Example of an identified magnetic bipole. The left
	panel shows magnetic field lines connecting the footpoints of
	the bipole as derived from the magnetofrictional
	simulation. Different field lines have different colors for
	easier identification. The middle panel shows the loop
	footpoints (red contours) in the NFI magnetogram and the right
	panel the corresponding intensity filtergram.  The
	observations were taken on November 3, 2010 at 01:34 UT.
	An animation showing the spatio-temporal evolution of
	the bipole is available in the online journal. The positive
	and negative polarity footpoints have lifetimes of $82.5$~min
	and $118.5$~min, respectively. Based on the magnetofrictional
	simulation, the footpoints are connected for about 40 
	minutes. The animation lasts 8 seconds and covers $\sim2$
	hours of solar time from November 3, 2010 at 01:17 UT.}
	\vspace*{1em} \label{fig4}
\end{figure*}

The time-dependent bottom boundary conditions needed to drive the
evolution of the magnetic field above the surface ($z>0$) are taken to
be
\begin{eqnarray}
-\frac{\partial A_x}{\partial t} &=& E_x, \\
-\frac{\partial A_y}{\partial t} &=& E_y,
\end{eqnarray}
where $E_x$ and $E_y$ are the horizontal components of the electric
field vector, computed as $E_{x}=v_{y} B_{z}$ and $E_{y}=-v_{x}
B_{z}$. Here, $v_{x}$ and $v_{y}$ represent the $x$ and $y$ components
of the horizontal velocity field, respectively. We used Local
Correlation Tracking \citep[LCT;][]{November} to determine $v_x$ and
$v_y$ from the proper motion of the magnetic features. The LCT
algorithm was applied to the NFI magnetogram sequence taking into
account all the pixels. However, in the resulting horizontal velocity
maps we set to zero the pixels with flux densities below 12~G
($3\sigma$). To avoid sharp changes in those pixels, we applied the
smoothing function
\begin{equation}
f_{v}=1-\frac{1}{1+e^{B_{z}^{\prime}-12}}+\frac{1}{1+e^{B_{z}^{\prime}+12}},
\end{equation}
where $B_{z}^{\prime}=B_{z}/1\mbox{G}$. An example of a horizontal
velocity map is presented in Figure~\ref{fig2}. Estimating the
electric field in this way does not, in general, give a pair $(E_{x},
E_{y})$ whose curl is equal to $-\partial B_{z}/\partial t$. Because
of this, we use the method presented in \cite{Mark2012} to
additionally solve for a correcting electric field to ensure the
boundary $B_{z}$ is consistent with the NFI magnetogram sequence. To
estimate the electric field, one could also use the method described
in \citeauthor{Kazachenkoetal2014} (\citeyear{Kazachenkoetal2014}; see
also \citealt{Fisheretal2015}, \citealt{Lummeetal2017},
\citealt{Priceetal2019}, \citealt{Hoeksemaetal2020}), but that would
require vector magnetograms which are not available here.

The computation was performed in a box of size $L_{x}\times
L_{y}\times L_{z}=61 \times 64 \times 21$~arcsec$^{3}$. This box extends sufficiently high into the solar atmosphere to prevent the upper boundary from influencing the results in the photosphere and the lower chromosphere. Open boundary conditions were imposed at the top of the computational box.

\section{Results}
\label{sect4}

\subsection{Magnetofrictional simulation}
A 3D rendering of the magnetic field lines resulting from the
magnetofrictional simulation of our data set is presented in
Figure~\ref{fig3}. The different colors of the field lines indicate
the cosine of the field inclination to the local vertical, from $+1$
(positive polarity, violet) to $-1$ (negative polarity, red). To our
knowledge, this is the first magnetofrictional simulation ever of a QS
region including a full supergranular cell. One can clearly see
small-scale, low-lying magnetic loops connecting patches of opposite
polarity, and open field lines associated with unipolar patches (the
mostly vertical red lines of negative polarity going away from the
depicted volume through the upper plane). The magnetic topology of the
region, together with its temporal evolution, holds the key to
separate bipolar features from unipolar features.

It is important to mention here  that the magnetofrictional simulation was tested for different FOV sizes. This was achieved by selecting different sizes of the Hinode/NFI FOV, as well as embedding the Hinode/NFI magnetograms into larger HMI magnetograms ($100\arcsec \times 100\arcsec$). The tests consistently produced open field lines extending up from the network regions above the supergranular cells. The field lines inside the supergranule seemed to be unaffected in the tests. Based on these results, we are confident
that the periodic boundaries do not affect the extrapolation of the field
lines inside the observed supergranular cell and that the general magnetic
morphology in the selected FOV is determined by the network
structures.

\begin{figure*}[t]
	\centering 
	\includegraphics[width=0.98\textwidth]{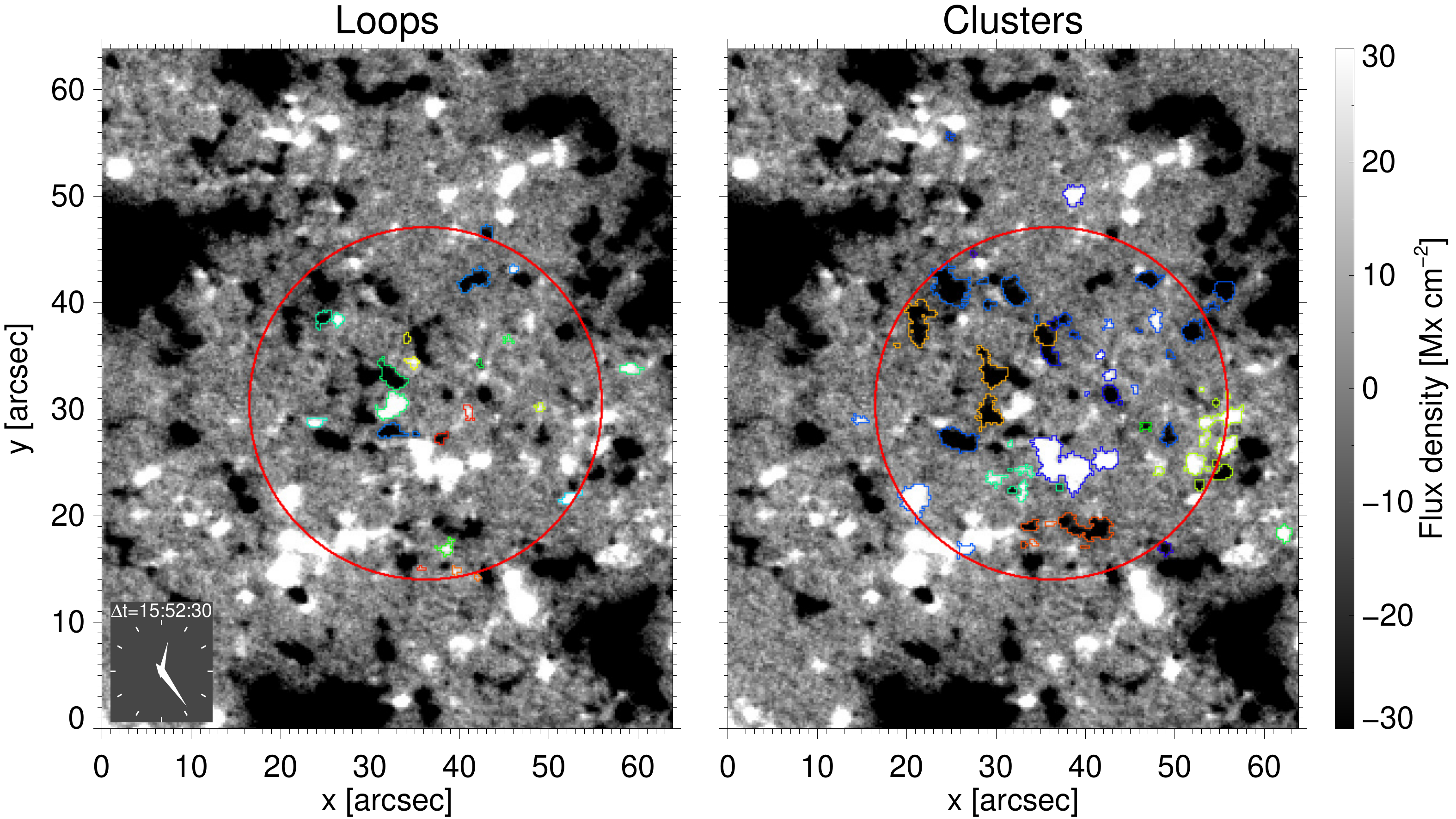}
	\caption{Magnetic loops (left) and clusters (right) detected
	in a Hinode magnetogram taken on November 3, 2010 at 00:24:04
	UT. Flux patches belonging to the same loops or clusters have
	the same contour colors. The magnetograms are saturated at
	$\pm30$~Mx~cm$^{-2}$. An animation of this figure is
	available in the online journal. It shows the spatio-temporal
	evolution of all detected loops and clusters. The animation
	lasts 96 seconds and covers $23.7$ hours of solar time from
	November 2, 2010 at 08:31 UT.}  
	\vspace*{1em} \label{fig5}
\end{figure*}

The quality of the results can be seen in the movie accompanying
Figure~\ref{fig4}, which shows an example of a magnetic bipole
detected in our magnetograms. The footpoints of the bipole emerged at
the edges of the same granule and separated from each other with
time. Using the magnetofrictional simulation we calculated the field
lines and confirmed that the two footpoints are magnetically
connected. The movie displays the complete history of the
footpoints. Eventually, they disappeared via fragmentation and
cancellation processes.

We would like to remind the reader that the simulation is not
sufficiently accurate for the weakest magnetic features whose signal
is close to the noise level. Such cases represent $\sim25$\%
of the detected loops but only $11$\% of the total bipolar flux, 
which is mainly determined by large loops and clusters.
 
\subsection{Tracking results}

The tracking of magnetic features in the selected region resulted in
$10661$ unique features, comprising $86732$ individual patches over
the whole sequence\footnote{Here, we distinguish between features and
patches. A feature is a physical object that can be followed from
birth to death and is seen as a series of individual magnetic patches
in consecutive time steps.}. A total of 8266 features appeared in
situ, 1102 and 1190 were formed by fragmentation of existing unipolar
and bipolar features, respectively, and 103 were present in the first
frame. The latter were excluded from the analysis because we do not
know their complete histories.

We classified the features appearing in situ as footpoints of magnetic
loops, cluster members, or unipolar features by considering all 6
aspects described in Sect.~\ref{sect33}. We found that 652 features
were loop footpoints and 1428 emerged in 155 clusters during 22 hours
of observations. This corresponds to $\sim$$8$\% and $\sim$$17$\% of
the total number of magnetic features that appeared in situ. The
remaining 6186 features ($75$\%) were unipolar.

Among all the features born in situ and by fragmentation, 31\% can be
classified as bipolar and 69\% as unipolar. However, the number of
flux patches associated with bipolar and unipolar features is not so
different (40686 vs 46046, or 47\% vs 53\%). This is because bipolar
features tend to live longer and therefore contribute more flux
patches per feature than their unipolar counterparts.

In the animation accompanying Figure~\ref{fig5} we show all the
magnetic loops (left panel) and clusters (right panel) detected in our
Hinode/NFI magnetograms. Flux patches belonging to the same loops or
clusters have the same colors. The red circles outline the area where
we look for bipolar features. The animation shows 22 hours of
observations during which we tracked magnetic bipoles and two hours
more to completely cover their lifetimes.

\subsection{Properties of unipolar and bipolar IN flux patches}
\label{properties}

\begin{figure*}[!t]
	\centering
	\resizebox{0.94\hsize}{!}{\includegraphics[width=0.9\textwidth]{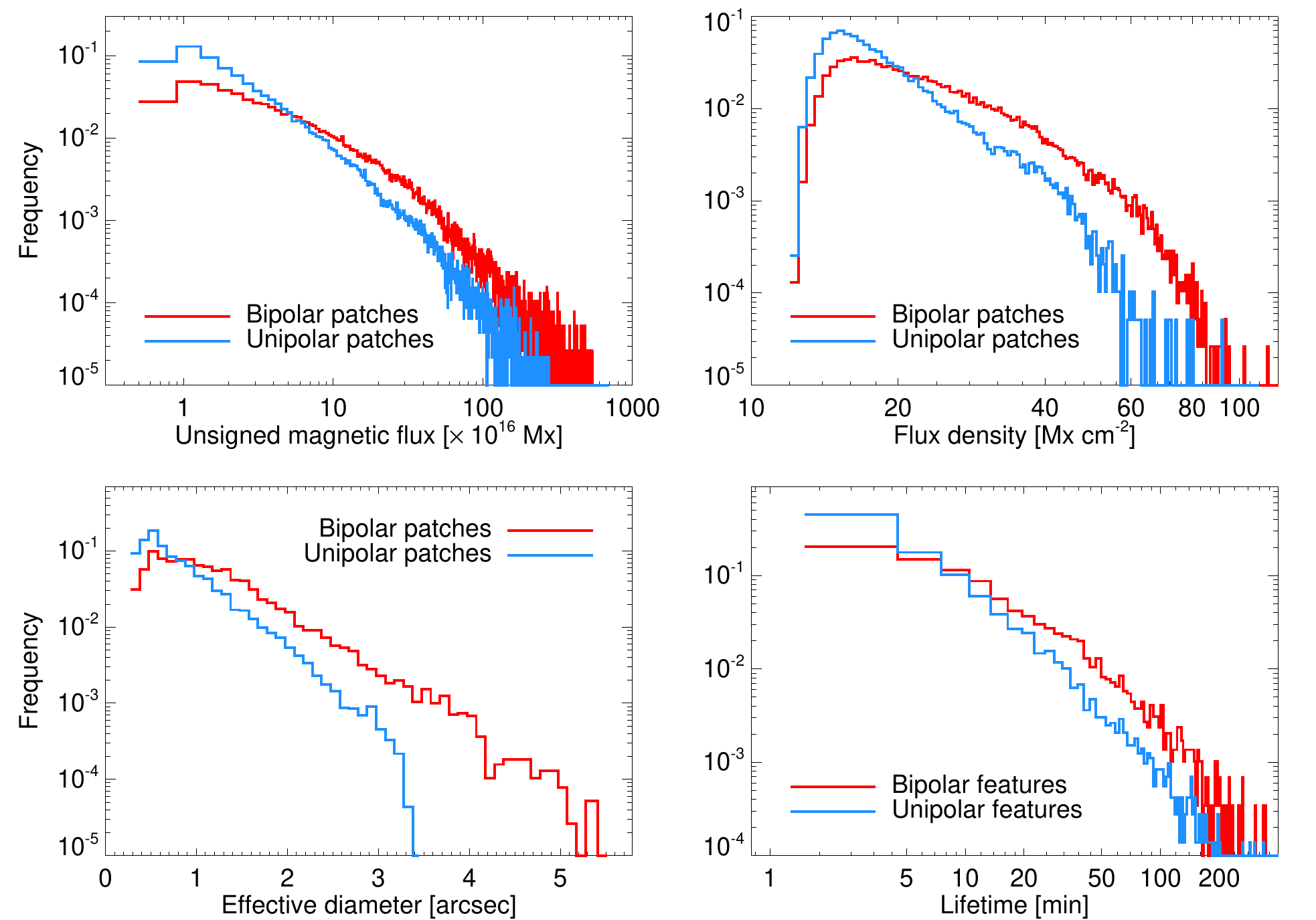}}
	\caption{Physical parameters of 46046 unipolar and 40686
	bipolar IN patches detected in this study. The upper panels
	show the unsigned magnetic flux (left) and flux density
	(right) distributions, while the bottom panels show the
	effective diameter (left) and feature lifetime (right)
	distributions. The blue and red solid lines represent unipolar
	and bipolar flux patches, respectively. The bin sizes used are
	$4 \times 10^{15}$~Mx (flux), 0.5 Mx~cm$^{ -2}$ (flux
	density), 0.1~arcsec (diameter), and 3 minutes (lifetime).}
    \vspace*{1em}
	\label{fig6}
\end{figure*}

To characterize the detected unipolar and bipolar magnetic patches, we
calculated their unsigned magnetic fluxes, magnetic flux densities,
sizes, and lifetimes. We define the unsigned magnetic flux of an individual 
patch as
\begin{equation}
	{\Phi}=\sum_{i=1}^{N}|\phi_{i}| \,{\rm d}A,
\end{equation}
where $N$ is the number of pixels in the patch, ${\rm d}A$ the area of
a pixel, and $\phi_{i}$ the magnetic flux density observed in pixel
$i$. We also compute the average magnetic flux density of the patch as
\begin{equation}
	\overline{\phi}=\frac{1}{N}\sum_{i=1}^{N}\phi_{i}.  
\end{equation}
This quantity provides an estimate of the longitudinal magnetic
field strength in the patch, assuming a filling factor equal to one.

The distributions of the various parameters are shown in
Fig.~\ref{fig6} separately for unipolar and bipolar patches. The
mean values over the two populations are given in Table~\ref{table1}.

As can be seen in the upper left panel of Figure \ref{fig6}, unipolar
patches span three decades in flux, from our detection limit of $6.5
\times 10^{15}$~Mx to about $3 \times 10^{18}$~Mx, with a mean value
of $9.2 \times 10^{16}$~Mx. The flux of bipolar patches spans from the
detection limit to approximately $5.4 \times 10^{18}$~Mx. Their mean
flux is $24.1 \times 10^{16}$~Mx, more than a factor $2.6$ larger than
that of unipolar patches. 

The lowest value of the flux density distribution is $12$~Mx~cm$^{-2}$
for both populations as a consequence of the $3\sigma$ threshold used
to identify the patches. The largest flux densities are about
$90$~Mx~cm$^{-2}$ and $115$~Mx~cm$^{-2}$ for unipolar and bipolar
patches, respectively. The mean flux densities of unipolar and bipolar
patches are $19.9$~Mx~cm$^{-2}$ and $25.7$~Mx~cm$^{-2}$,
respectively. Therefore, bipolar patches tend to be stronger. The
strongest flux concentrations correspond to those appearing in
clusters.

Table~\ref{table1} shows that bipolar patches are in nearly perfect
flux balance, with positive patches accounting for 51\% of the total
bipolar flux observed in the cell. By contrast, unipolar patches
exhibit a certain amount of flux imbalance: about 62\% of the total
unipolar flux is negative due to the surplus of negative patches that
is detected. Interestingly, also the supergranular cell is dominated
by negative polarity fields. For this reason, the possibility exists
that a fraction of the negative unipolar patches are actually formed
by accumulation of background flux from the supergranule itself and
not by a more flux-balanced mechanism.

The mean effective diameter is about $0.8$~arcsec for unipolar patches and about $1.1$~arcsec for bipolar patches. Here the effective diameter is defined as that of a circular structure with the
same area as the identified magnetic patch. The largest unipolar and
bipolar patches in our data set have effective diameters of $3.3$ and
$5.4$~arcsec, respectively. Thus, most bipolar patches are bigger than
unipolar patches, although some of them can be very small too.

\begin{deluxetable}{lcc}
	\tablecolumns{3} \tablewidth{\columnwidth}
	\tablecaption{Physical parameters of the detected unipolar and
	bipolar IN patches} \tablehead{ \colhead{} &
	\colhead{Unipolar} & \colhead{Bipolar}} 
	\startdata 
	Total number of features & 7288 & 3270 \\ 
	Total number of patches & 46046 & 40686 \\ 
		\hspace{1em} Number of positive patches & 19486 & 19173\\ 
		\hspace{1em} Number of negative patches & 26560 & 21513 \\ 
	Mean unsigned flux [$10^{16}$ Mx] & 9.2 & 24.1\\ 
		\hspace{1em} positive patches [$10^{16}$ Mx] & 8.3 & 25.9 \\ 
		\hspace{1em} negative patches [$10^{16}$ Mx] & 9.9 & 22.4 \\ 
	Total unsigned flux carried by   &  & \\ 
		\hspace{1em} positive patches [$10^{21}$ Mx] & 1.6 & 5.0 \\ 
		\hspace{1em} negative patches [$10^{21}$ Mx] & $2.6$ & 4.8 \\ 
	Mean flux density [Mx cm$^{-2}$] & 19.9 & 25.7 \\ 
	Mean effective diameter [arcsec] & 0.8 & 1.1 \\ 
	Mean lifetime [min] & 10.7 & 23.4 \enddata 
\end{deluxetable}
\label{table1}

Similarly to \cite{Gosic2014, Gosic2016}, we consider a feature to
live from the frame in which it receives a unique label until the
moment it loses its label. In our magnetogram sequence, the lifetimes
range from 1.5~min up to $\sim330$~min (unipolar features) and
$\sim345$~min (bipolar features). The lower limit is set by the
detection threshold. On average, bipolar features live significantly
longer than unipolar features ($23.4$ vs $10.7$~min). This is because
they tend to be larger, so they have a better chance to survive
interactions with other magnetic features and require more time to
disperse. If we consider bipolar loops and clusters to live
from the moment when the first footpoint or member appears until all
of them disappear, then their average lifetime is even longer
($62$~min).

Magnetic features that have long lifetimes usually appear in
clusters. They undergo multiple mergings which help them accumulate
enough flux to withstand convective dragging, fragmentation and
partial cancellation over longer periods of time. The red arrow in
Figure \ref{fig5} indicates one such feature. However, long lifetimes
are not only associated with bipolar features. Although less common,
unipolar features can also live for 4-5 hours when they find
themselves in the right environment, i.e., when they are surrounded
mostly by same-polarity features and have enough time for merging
processes to occur multiple times.

Long-lived features are strong magnetic flux concentrations 
and may have a substantial impact on the QS chromosphere. This could
be realized, for example, through reconnection between the preexisting
ambient fields and the IN flux concentrations emerging into the
chromosphere \citep{Gosicetal2021}. It is also possible that the
long-lived IN features may affect the chromospheric velocity field and
the low-lying canopy that extends above supergranular cells
\citep{Robustinietal2019}.

\begin{figure*}[t]
	\centering \includegraphics[width=0.88\textwidth]{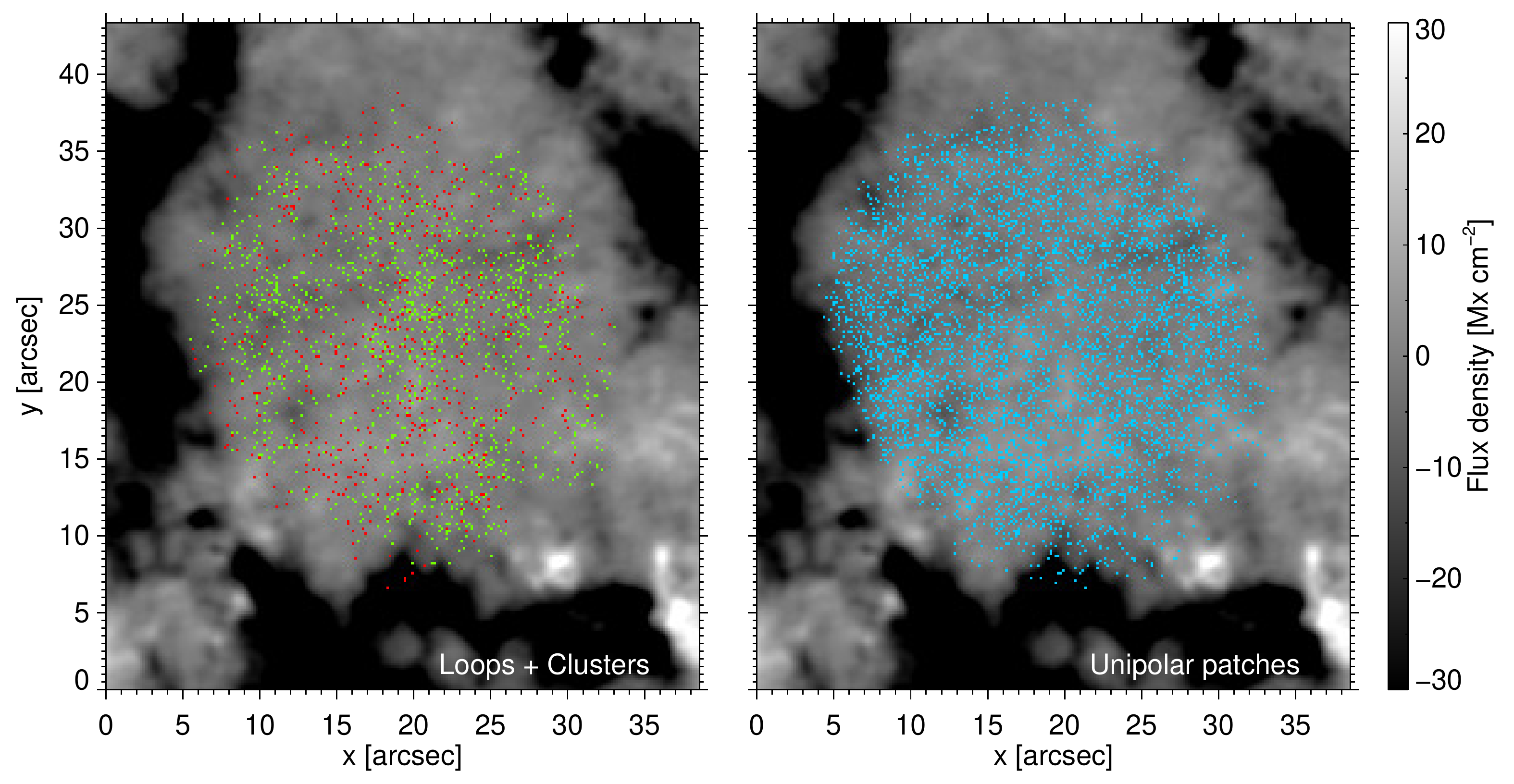}
	\caption{Location of appearance of bipolar (left) and unipolar
	(right) features inside the supergranular cell. Green dots
	represent flux features emerging in clusters while the
	footpoints of magnetic loops are shown with red dots. Blue
	dots mark where unipolar features appear. The background image
	is the mean magnetogram averaged over the 22 hours duration of
	the time sequence.}  \label{fig7}
\end{figure*}

\begin{figure*}[t]
	\centering \includegraphics[width=0.99\textwidth, bb=30 0 887
	283]{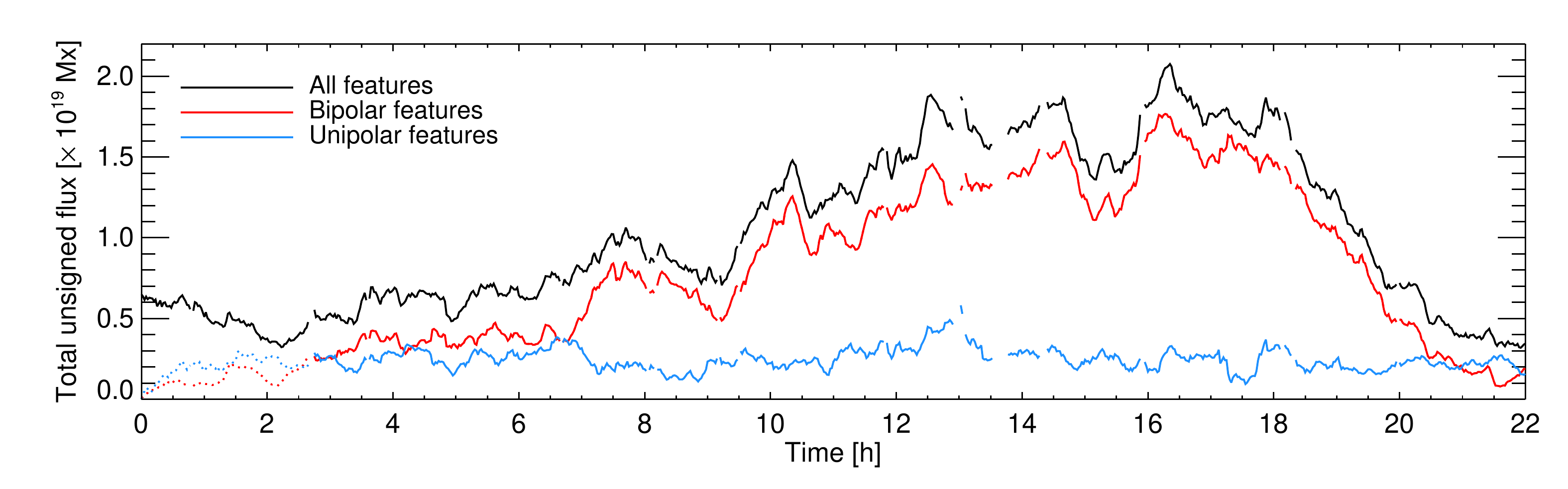} \caption{Temporal evolution of the total
	instantaneous unsigned flux carried by the unipolar (blue) and
	bipolar (red) features that originated within the selected
	circular region, along with their sum (black). The dotted
	lines mark the period without reliable data due to the
	existence of features at the beginning of the sequence whose
	modes of appearance are not known.}  
   \label{fig8}
   \vspace*{1em}
\end{figure*}

\begin{figure*}[t]
	\centering \includegraphics[width=0.99\textwidth, bb=30 0 887
	283]{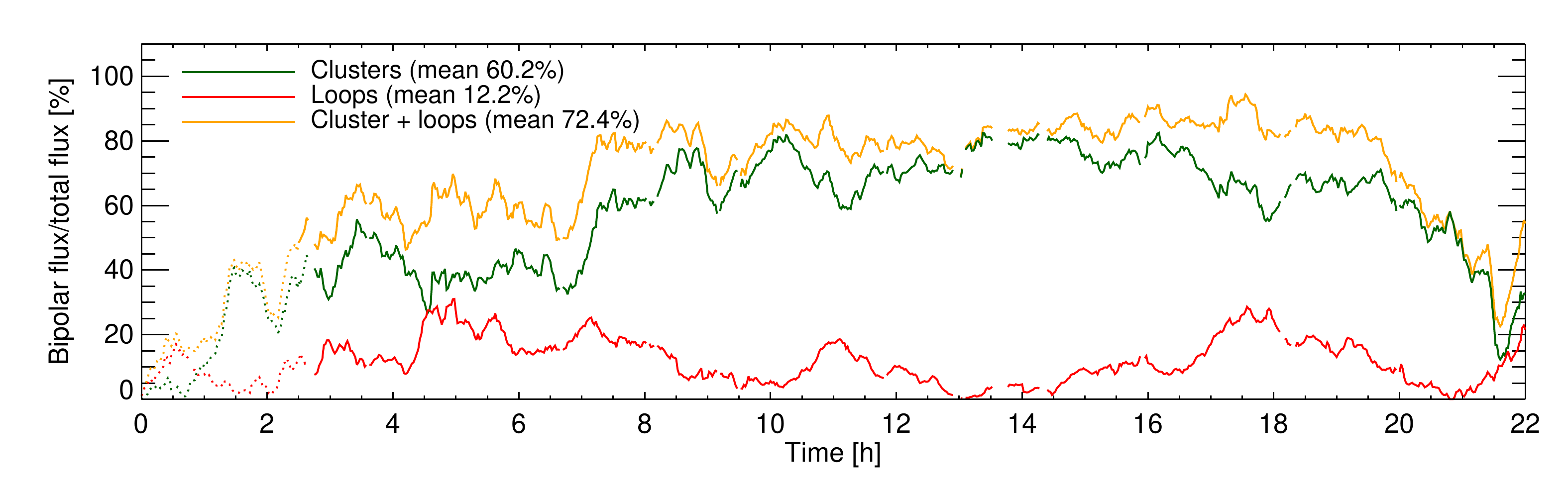} 
	\caption{Fraction of the total instantaneous 
	unsigned flux within the selected supergranular region in the
	form of bipolar features. The green and red lines correspond
	to clusters and loops, respectively, while the orange line
	shows their sum. The dotted lines mark incomplete data at the
	beginning of the sequence.}  
	\label{fig9} 
\end{figure*}

\subsection{Spatial distribution}

Our observations reveal that bipolar features emerge everywhere in the
selected region. In Figure~\ref{fig7} we plot the locations of
appearance of all bipolar and unipolar features. As can be seen, loops
and clusters are more or less uniformly distributed inside the circle
and do not show any preferred emergence location after 22 hours of
observations. Unipolar features also appear uniformly across the
supergranule. We would like to remind the reader that the appearance
locations represent the flux-weighted centers of the magnetic patches
at birth. Therefore, the actual area occupied by them is larger and
covers the entire cell. Indeed, over a period of 22~h we do not see
any dead calm area of the type detected by \cite{MartinezGonzalezetal2012} 
in SUNRISE/IMaX magnetogram sequences of about 30~min duration.

\subsection{Unipolar and bipolar flux budget}

\begin{figure*}[t]
	\centering
	\includegraphics[width=0.99\textwidth, bb=30 0 887 283]{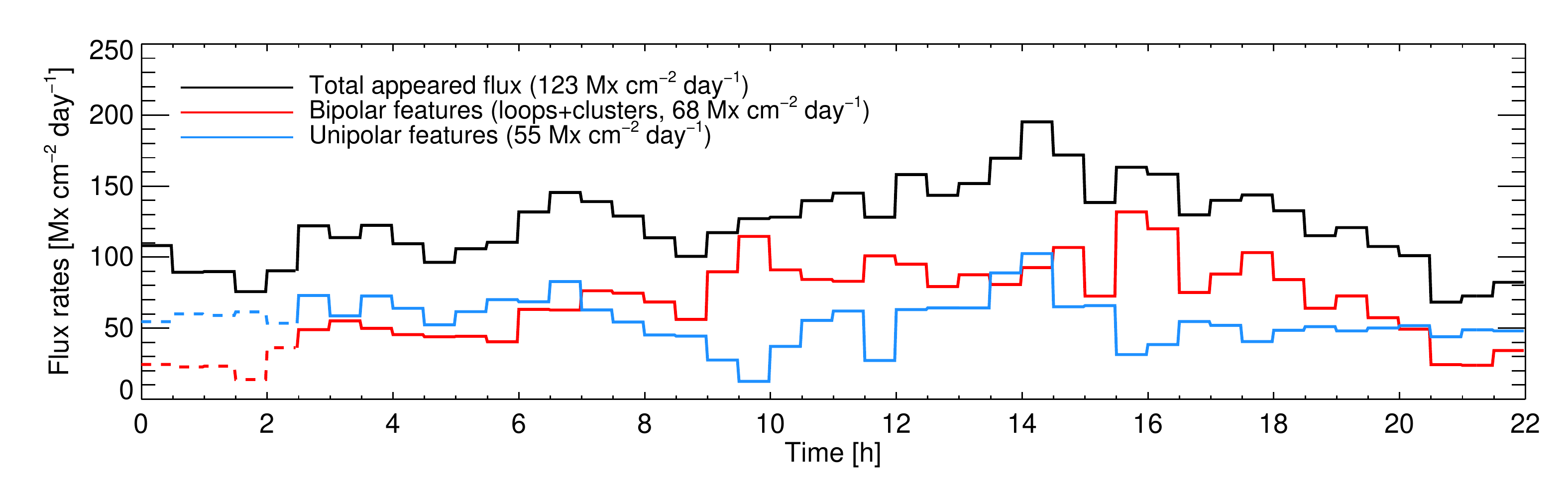} 
	\caption{Appearance rates of bipolar and unipolar flux in the IN.
		The red lines shows the flux brought to the surface by magnetic
		loops and clusters at birth and during their evolution, 
		while the blue line represents unipolar features. The total 
		appearance rate is shown in black. To improve the statistics, 
		bins of $0.5$~h are used.}
	\label{fig10}
	\vspace*{1em}
\end{figure*}

The total instantaneous unsigned fluxes of unipolar and bipolar IN
features are shown in Figure~\ref{fig8} (blue and red curves,
respectively). The first three hours are not reliable since we do not
know the modes of appearance of the features that were visible in the
first frame. It was necessary to wait 3~h for all those features to
disappear. We can see here that the flux carried by unipolar features
is practically constant with time. On the other hand, the flux brought
by loops and clusters shows strong variations, especially when large
clusters emerge. Bipolar features experience an intrinsic growth in
flux and size over their lifetime, and this partly explains the
positive excursions associated with the appearance of
clusters. Moreover, the two curves are affected by interactions
between features, which mix the unipolar and bipolar fluxes. For
example, when a small unipolar feature merges with a strong feature
from a cluster and cannot be tracked any longer, the flux of the
smaller feature is added to the flux of the cluster, i.e., the total
bipolar flux increases at the expense of the unipolar flux. However,
this is balanced to some extent as the same process happens in the
opposite direction as well. Finally, the curves are also affected by
mergings and cancellations of IN patches with NE elements, which cause
drops in the total amount of flux (stronger for bipolar features,
as they tend to be bigger and carry more flux per feature).

The contribution of bipolar features to the total instantaneous flux
of this supergranular cell is very significant, as can be seen in
Figure~\ref{fig9}. After the initial three hours, the flux carried by
clusters represents about $60$\% of the total detected flux, while the
loops contain $12$\% of the detected flux. Together, clusters and
loops account for $\sim$$72$\% of the IN flux, with temporal
variations from $50$\% to $95$\%. The rest of the flux in this
supergranular cell is in the form of unipolar features. Interestingly,
close to the end of the sequence there is a 30-minute interval when
the bipolar flux accounts for only $\sim$$20$\% of the total flux. This
period coincides with a lack of large bipolar features.

\subsection{Unipolar and bipolar flux appearance rate}

Figure~\ref{fig10} shows the flux appearance rate of bipolar (loops
and clusters) and unipolar features as a function of time. The data
have been binned over 30 minutes. Each bin represents the sum of the
flux brought to the surface by new features and the flux gained by
already existing features during that period of time (for more
details, see \citealt{Gosic2016}). The total flux appearance rate 
in the central part of the supergranular cell is
$123$~Mx~cm$^{-2}$~day$^{-1}$, in agreement with the value reported by
\cite{Gosic2016}. Bipolar flux emerges on the surface at a rate of
$68$~Mx~cm$^{-2}$~day$^{-1}$, or 55\% of the total rate. The remaining
$55$~Mx~cm$^{-2}$~day$^{-1}$ (45\% of the total rate) is due to
unipolar features.

In Figure~\ref{fig10}, all the curves show temporal variations. In the
case of loops and clusters, the variations are larger when strong
bipolar features emerge, for example around 9, 11 and 16~h. Unipolar
fields are more stable: their appearance rates are almost constant in
the first $8$ hours and in the last 6 hours. However, they fluctuate
when clusters emerge. It may happen that in such periods we detect
fewer unipolar patches because the strong bipolar patches occupy a
larger surface area. The appearance rate of bipolar flux in the first
three hours is most likely underestimated, explaining why it is so
low. The reason is that we cannot classify as unipolar or bipolar the
magnetic features that were visible in the first frame. These features
and their fragments were discarded from the analysis during the first
three hours (the time they needed to completely leave the selected
region), so any flux gain they might have experienced during
that time was not counted. This primarily affects large magnetic
features such as those that normally emerge in bipolar form. In any
case, the black curve shows that the total flux appearance rate was
normal during the first three hours, even if this is not reflected in
the other curves.

Finally, we should keep in mind that unipolar features in our data set
are smaller and contain less flux than bipolar features. Many of them
have signals that fluctuate around the threshold level. Because of
this, they may disappear and reappear frequently in the
magnetograms. Thus, even though we correct for reappearing events as
in \cite{Gosic2016}, unipolar features do not necessarily represent
new flux brought to the surface. For that reason, their appearance 
rate should be considered as an upper limit.

\section{Discussion and Conclusions}
\label{sect5}

In this paper, we studied the appearance modes and the temporal
evolution of IN magnetic features inside a supergranular cell. We used
a high-resolution, high-sensitivity, long-duration Hinode/NFI
magnetogram sequence taken on 2010 November 2--3. We tracked flux
features inside the supergranule and for the first time we employed
magnetofrictional simulations to identify unipolar and bipolar
features in the cell interior. We then determined how and where these
fields appear and calculated the total unsigned flux they bring to
the surface. Our findings can be summarized as follows:

\begin{enumerate}
	\item Bipolar features (magnetic loops and flux clusters)
	appear more or less uniformly inside the supergranular
	cell. This is at odds with \cite{MartinezGonzalezetal2012} and
	\cite{Stangalini2014}.  Perhaps the reason is the
	shorter duration of the time sequences they used, which may
	have resulted in insufficient statistics (particularly for
	clusters). In our data set, on supergranular time scales all
	the observed area eventually undergoes emergence
	processes. Thus, the magnetic voids reported in the literature
	are short-lived patterns (existing on granular/mesogranular
	time scales only) or do not occur in the cell we have
	studied. It is important to mention that there are indeed
	periods where less magnetic flux is brought to the solar
	surface, which can be easily identified in Figures \ref{fig1},
	\ref{fig8}, \ref{fig9}, and \ref{fig10}. However, even during
	those periods magnetic patches appear evenly distributed
	across the supergranule, as can be seen in Figure 5.13 of
	\cite{milan}.

	\item Bipolar features emerge in the interior of the
	supergranular cell at a rate of
	68~Mx~cm$^{-2}$~day$^{-1}$. This is new IN flux coming most
	likely from below the surface. It accounts for 55\% of the
	total flux appearance rate.

	\item Unipolar features appear at a rate of
	55~Mx~cm$^{-2}$~day$^{-1}$, or 45\% of the total flux
	appearance rate.

	\item On average, bipolar features contain about 72\% of the
	total instantaneous IN flux. This value is observed to
	fluctuate between 50\% and 95\% with time. It drops to 20\%
	for only about 30 minutes close to the end of the
	sequence. The fraction of bipolar instantaneous flux is larger
	than the bipolar flux appearance rate due to two processes:
	the intrinsic growth of bipolar features in flux and size
	during the early phases of emergence, and the transfer of flux
	that happens when (usually small) unipolar features merge with
	bipolar features and disappear as individual elements.
\end{enumerate}

Our results lend support to the idea that there are two distinct
populations of IN flux concentrations inside supergranular cells. One
group consists of unipolar features while the second group are bipolar
features. Unipolar flux concentrations are probably formed by
coalescence of background flux, as pointed out by \cite{Lamb2008} and
also suggested by the analysis presented by \cite{Gosic2016}.

\begin{figure}[!t]
	\centering \includegraphics[width=0.45\textwidth, bb=20 0 453
	311]{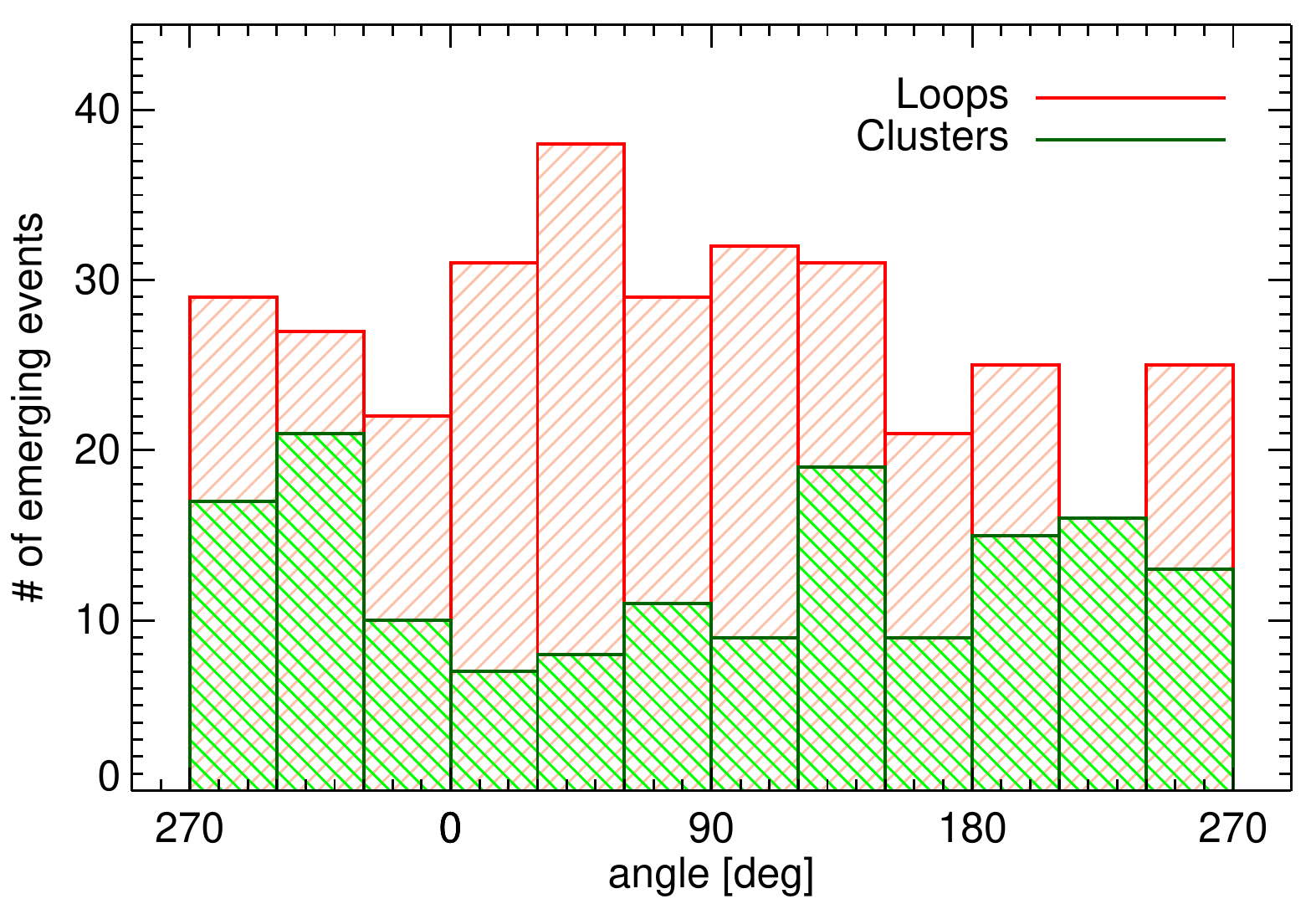} 
	\caption{Orientation of the magnetic axes of
		bipolar features emerging in the IN with respect to solar
		East-West. The axis is defined as the vector connecting the
		negative and positive footpoints of loops (red) or the
		flux-weighted centers of negative and positive flux patches in
		clusters (green).  The angles increase counterclockwise. An
		angle of $0^{\rm o}$ represents a magnetic bipole parallel to
		the solar equator, pointing West. The data are binned in
		$30^{\circ}$ intervals.}  
	\label{fig11}
\end{figure}

Bipolar features may be the signature of local dynamo action or a
result of the global dynamo. The second possibility is perhaps less
probable, given that the two footpoints of the loops and the
opposite-polarity centers of the clusters do not seem to show a
systematic orientation upon appearance. This is illustrated in Figure~\ref{fig11}. Thus, either they are not oriented in any preferred direction (favoring the local dynamo scenario) or the
orientation they adopt is largely determined by convective flows
during the emergence process (not by the orientation they have deep in
the convection zone).

The results presented in this work partially agree with those based on
SUNRISE/IMaX data \citep{Anusha2017, Smithaetal2017}. We confirm the
IMaX result that unipolar features are more numerous than bipolar
features. However, in our magnetograms bipolar features account for
55\% of the flux appearance rate, as opposed to only $\sim$9\% in the
IMaX data \citep{Smithaetal2017}. We should keep in mind that the IMaX
observations have shorter duration than the NFI sequence used here,
i.e., they cover only a very small fraction of the typical
supergranular time scales. Statistical fluctuations cannot be ruled
out in this case. In particular, it is likely that IMaX did not
capture clusters and/or large magnetic loops, which would
significantly decrease the contribution of bipolar features to the
flux appearance rate. Also, the particular ability of the tracking
algorithms to recognize magnetic elements as unipolar or bipolar
features may have caused additional differences among the reported
rates.

Another source of discrepancy is the total flux appearance rate
itself. The values inferred by \cite{Smithaetal2017} are an order of
magnitude larger than ours. The difference may be due to the higher
sensitivity and spatial resolution of the IMaX observations, together
with the use of a lower detection threshold ($2\sigma$ vs $3\sigma$)
and no minimum lifetime for features to be included in the
analysis. This led to the detection of much weaker flux features in
the IMaX magnetograms as compared with the NFI observations ($9 \times
10^{14}$~Mx vs $5
\times 10^{15}$~Mx) and many more features per frame, which may 
well explain the different total flux appearance rates. Finally, one
should not forget that the spectral lines observed by Hinode and IMaX
are different -- they have different magnetic sensitivities and sample
different atmospheric layers. The Hinode/NFI line is formed in the
mid/upper photosphere and therefore can be expected to yield fewer
features (because most of them lie in the lower photosphere) and
weaker fluxes (due to the general reduction of the magnetic field with
height). This might also partially explain the different results
obtained from Hinode/NFI and SUNRISE/IMaX.

To better understand the magnetism of the QS, we need to
investigate how different methods and instruments may affect the
analysis of the flux appearance modes in the QS. However, it is also
clear that we will need measurements at higher resolution and
sensitivity to capture the weakest magnetic fields of the solar
internetwork. This will soon be made possible by a new generation of
telescopes, particularly the Daniel K.\ Inouye Solar Telescope
\citep{Rimmeleetal2020}.

\begin{acknowledgments}
The data used here were acquired in the framework of the
Hinode Operation Plan 151 {\em ``Flux replacement in the
solar network and internetwork.''} We thank the Hinode Chief
Observers for the efforts they made to accommodate our demanding
observations. Hinode is a Japanese mission developed and
launched by ISAS/JAXA, with NAOJ as a domestic partner and NASA and
STFC (UK) as international partners. It is operated by these agencies
in co-operation with ESA and NSC (Norway). MG acknowledges a JAE-Pre
fellowship granted by Agencia Estatal Consejo Superior de
Investigaciones Cient\'{\i}ficas (CSIC) toward the completion of a PhD
degree. This work has been funded by the State Agency for Research of
the Spanish Ministerio de Ciencia e Innovaci\'on through grant
RTI2018-096886-B-C5 (including FEDER funds) and through the ``Center
of Excellence Severo Ochoa" award to the Instituto de Astrof\'{\i}sica
de Andaluc\'{\i}a (SEV-2017-0709). NASA supported this work through
contract NNM07AA01C (Solar-B (Hinode) Focal Plane Package Phase
E). Use of NASA's Astrophysical Data System is gratefully
acknowledged.
\end{acknowledgments}

\end{document}